\documentclass[prd,reprint,showpacs,showkeys]{revtex4-1}
\usepackage{amssymb}
\usepackage{amsmath}
\usepackage{graphicx}
\usepackage[font={footnotesize,it}]{caption}

\pdfoutput=1

\begin{document}

\title{Thin-shell wormholes from the regular Hayward black hole}
\author{M. Halilsoy}
\email{mustafa.halilsoy@emu.edu.tr}
\author{A. Ovgun}
\email{ali.ovgun@emu.edu.tr}
\author{S. Habib Mazharimousavi}
\email{habib.mazhari@emu.edu.tr}
\affiliation{Department of Physics, Eastern Mediterranean University, G. Magusa, north
Cyprus, Mersin 10 - Turkey}
\date{\today }

\begin{abstract}
We revisit the regular black hole found by Hayward in $4-$dimensional
static, spherically symmetric spacetime. To find a possible source for such
a spacetime we resort to the non-linear electrodynamics in general
relativity. It is found that a magnetic field within this context gives rise
to the regular Hayward black hole. By employing such a regular black hole we
construct a thin-shell wormhole for the case of various equations of state
on the shell. We abbreviate a general equation of state by $p=\psi \left(
\sigma \right) $ where $p$ is the surface pressure which is the function of
the mass density ($\sigma $). In particular, a linear, logarithmic,
Chaplygin, etc. forms of equations of state are considered. In each case we
study the stability of the thin-shell against linear perturbations. We plot
the stability regions by tuning the parameters of the theory. It is observed
that the role of the Hayward parameter is to make the TSW more stable.
Perturbations of the throat with small velocity condition is also studied.
The matter of our TSWs, however, remains to be exotic.
\end{abstract}

\maketitle

\section{Introduction}

Thin-shell wormholes (TSWs) constitute one of the wormhole classes in which
the exotic matter is confined on a hypersurface and therefore can be
minimized \cite{1} (The $d-$dimensional thin-shell wormholes is considered
in \cite{Dias} and with a cosmological constant is studied in \cite{Lobo}).
Finding a physical (i.e. non-exotic) source to wormholes of any kind remains
ever a challenging problem in Einstein's general relativity. In this regard
we must add that modified theories of gravity presents more alternatives
with their extra degrees of freedom. We recall, however, that each modified
theory partly cures while partly adds its own complications. Staying within
Einstein's general relativity and finding remedies seems to be the prominent
approach provided the proper spacetimes are employed. An interesting class
of spacetimes that may serve the purpose is the spacetimes of regular black
holes.

\textit{Our motivation for choosing a regular black hole in wormhole
construction can be justified by the fact that a regular system can be
established from a finite energy.} \textit{In the high energy collision
experiments for instance, formation of such regular objects are more tenable.%
} Such a black hole was discovered first by Bardeen and came to be known as
Bardeen black hole \cite{2,Ayon}. Ayon-Beato and Garcia in their Letter \cite%
{Ayon} introduced a non-linear electric field source for the Bardeen black
hole. Bronnikov, later on, showed that the regular "electric" black holes
e.g., the one considered by Ayon-Beato and A. Garcia, are not quite correct
solutions to the field equations because in these solutions the
electromagnetic Lagrangian is inevitably different in different parts of
space. On the contrary, quite correct solutions of this kind (and even with
the same metric) can be readily obtained with a magnetic field (since in NED
there is no such duality as in the linear Maxwell theory). All this is
described in detail in \cite{Bronnikov}. A similar type of black hole
solution was given by Hayward \cite{3} which provides the main motivation
and fuel to the present study. This particular black hole solution has
well-defined asymptotic limits, namely it is Schwarzschild for $r\rightarrow
\infty $ and de-Sitter for $r\rightarrow 0.$ In order to make a better
account of the Hayward black hole we attempt first to explore its physical
source. For this reason we search for the non-linear electrodynamics (NED)
and find that a magnetic field within this theory accounts for such a
source. Note that every NED doesn't admit a linear Maxwell limit and indeed
this is precisely the case that we face in the present problem. In other
words, if our NED model did have a Maxwell limit then the Hayward spacetime
should coincide with the Reissner-Nordstr\"{o}m (RN) limit. Such a limit
doesn't exist in the present problem. Once we fix our bulk spacetime the
next step is to locate the thin-shell which must lie outside the event
horizon of the black hole. The surface energy-momentum tensor on the shell
must satisfy the Israel junction conditions \cite{4}. As the Equation of
State (EoS) for the energy-momentum on the shell we choose different models
which are abbreviated by $p=\psi \left( \sigma \right) $. Here $p$ stands
for the surface pressure, $\sigma $ is the mass (energy) density and $\psi
\left( \sigma \right) $ is a function of $\sigma .$ We consider the
following cases: i) Linear gas (LG) \cite{5}, where $\psi \left( \sigma
\right) $ is a linear function of $\sigma .$ ii) Chaplygin gas (CG) \cite{6}%
, where $\psi \left( \sigma \right) \sim \frac{1}{\sigma }.$ iii)
Generalized Chaplygin gas (GCG) \cite{7}, where $\psi \left( \sigma \right)
\sim \frac{1}{\sigma ^{\nu }}$ ($\nu =$ constant). iv) Modified Generalized
Chaplygin Gas (MGCG) \cite{8}, where $\psi \left( \sigma \right) \sim $%
LG+GCG. v) and Logarithmic Gas (LogG), where $\psi \left( \sigma \right)
\sim \ln \left\vert \sigma \right\vert .$ For each of the case we plot the
second derivative of the derived potential function $V^{\prime \prime
}\left( a_{0}\right) $, where $a_{0}$ stands for the equilibrium point. The
region that the second derivative is positive (i.e. $V^{\prime \prime
}\left( a_{0}\right) >0$) yields the regions of stability which are all
depicted in figures. This summarizes our strategy that we adopt in the
present paper for the stability of the thin shell wormholes constructed from
the Hayward black hole.

Organization of the paper is as follows. Section II reviews the Hayward
black hole and determines a Lagrangian for it. Derivation of the stability
condition is carried out in section III. Particular examples of equations of
state follow in section IV. Small velocity perturbations is the subject of
section V. The paper ends with the Conclusion in section VI.

\section{Regular Hayward black hole}

The spherically symmetric static Hayward nonsingular black hole introduced
in \cite{3} is given by the following line element%
\begin{multline}
ds^{2}=-\left( 1-\frac{2mr^{2}}{r^{3}+2ml^{2}}\right) dt^{2}+ \\
\left( 1-\frac{2mr^{2}}{r^{3}+2ml^{2}}\right) ^{-1}dr^{2}+r^{2}d\Omega ^{2}
\end{multline}%
in which $m$ and $l$ are two free parameters and 
\begin{equation}
d\Omega ^{2}=d\theta ^{2}+\sin ^{2}\theta d\phi ^{2}.
\end{equation}%
The metric function of this black hole $f\left( r\right) =\left( 1-\frac{%
2mr^{2}}{r^{3}+2ml^{2}}\right) $ at large $r$ behaves as%
\begin{equation}
\lim_{r\rightarrow \infty }f\left( r\right) \rightarrow 1-\frac{2m}{r}+%
\mathcal{O}\left( \frac{1}{r^{4}}\right)
\end{equation}%
while at small $r$ 
\begin{equation}
\lim_{r\rightarrow 0}f\left( r\right) \rightarrow 1-\frac{r^{2}}{l^{2}}+%
\mathcal{O}\left( r^{5}\right) .
\end{equation}%
From the asymptotic form of the metric function at small and large $r$ one
observes that the Hayward nonsingular black hole is a de-Sitter black hole
for small $r$ and Schwarzschild spacetime for large $r$. The curvature
scalars are all finite at $r=0$ \cite{9}. The Hayward black hole admits
event horizon which is the largest real root of the following equation%
\begin{equation}
r^{3}-2mr^{2}+2ml^{2}=0.
\end{equation}%
Setting $r=m\rho $ and $l=m\lambda $ this becomes%
\begin{equation}
\rho ^{3}-2\rho ^{2}+2\lambda ^{2}=0
\end{equation}%
which admits no horizon (regular particle solution) for $\lambda ^{2}>\frac{%
16}{27},$ single horizon (regular extremal black hole) for $\lambda ^{2}=%
\frac{16}{27},$ and double horizons (regular black hole with two horizons)
for $\lambda ^{2}<\frac{16}{27}$. Therefore the important parameter is the
ratio $\frac{l}{m}$ with critical ratio at $\left( \frac{l}{m}\right)
_{crit.}=\frac{4}{3\sqrt{3}}$ but not $l$ and $m$ separately. This suggests
to set $m=1$ in the sequel without loss of generality i.e., $f\left(
r\right) =1-\frac{2r^{2}}{r^{3}+2l^{2}}.$ Accordingly for $l^{2}<\frac{16}{27%
}$ the event horizon is given by%
\begin{equation}
r_{h}=\frac{1}{3}\left( \sqrt[3]{\Delta }+\frac{4}{\sqrt[3]{\Delta }}%
+2\right)
\end{equation}%
in which $\Delta =8-27l^{2}+3\sqrt{27l^{2}\left( 3l^{2}-2\right) }.$ For the
case of extremal black hole i.e. $l^{2}=\frac{16}{27}$ the single horizon
occurs at $r_{h}=\frac{4}{3}.$ For the case $l^{2}\leq \frac{16}{27}$ the
standard Hawking temperature at the event horizon is given by%
\begin{equation}
T_{H}=\frac{f^{\prime }\left( r_{h}\right) }{4\pi }=\frac{1}{4\pi }\left( 
\frac{3}{2}-\frac{2}{r_{h}}\right)
\end{equation}%
which clearly for $l^{2}=\frac{16}{27}$ vanishes and for $l^{2}<\frac{16}{27}
$ is positive (One must note that $r_{h}\geq \frac{4}{3}$). Considering the
standard definition for the entropy of the black hole $S=\frac{\mathcal{A}}{4%
}$ in which for $\mathcal{A}=4\pi r_{h}^{2}$ one finds the heat capacity of
the black hole defined by%
\begin{equation}
C_{l}=\left( T_{H}\frac{\partial S}{\partial T_{H}}\right) _{l}
\end{equation}%
and determined as%
\begin{equation}
C_{l}=4\pi r_{h}^{3}\left( \frac{3}{2}-\frac{2}{r_{h}}\right)
\end{equation}%
which is clearly non-negative. The fact that $C_{l}>0$ shows that
thermodynamically the black hole is stable.

\subsection{Magnetic monopole field as a source for the Hayward black hole}

We consider the action 
\begin{equation}
\mathcal{I}=\frac{1}{16\pi }\int d^{4}x\sqrt{-g}\left( R-\mathcal{L}\left(
F\right) \right)
\end{equation}%
in which $R$ is the Ricci scalar and 
\begin{equation}
\mathcal{L}\left( F\right) =-\frac{24m^{2}l^{2}}{\left[ \left( \frac{2P^{2}}{%
F}\right) ^{3/4}+2ml^{2}\right] ^{2}}=-\frac{6}{l^{2}\left[ 1+\left( \frac{%
\beta }{F}\right) ^{3/4}\right] ^{2}}
\end{equation}%
is the nonlinear magnetic field Lagrangian density with $F=F_{\mu \nu
}F^{\mu \nu },$ the Maxwell invariant with $l$ and $\beta $ two constant
positive parameters. Let us note that the subsequent analysis will fix $%
\beta $ in terms of the other parameters. The magnetic field two form is
given by%
\begin{equation}
\mathbf{F}=P\sin ^{2}\theta d\theta \wedge d\phi
\end{equation}%
in which $P$ stands for the magnetic monopole charge. This field form
together with the line element (1) imply%
\begin{equation}
F=\frac{2P^{2}}{r^{4}}.
\end{equation}%
The Einstein-NED field equations are ($8\pi G=c=1$) 
\begin{equation}
G_{\mu }^{\nu }=T_{\mu }^{\nu }
\end{equation}%
in which 
\begin{equation}
T_{\mu }^{\nu }=-\frac{1}{2}\left( \mathcal{L}\delta _{\mu }^{\nu }-4F_{\mu
\lambda }F^{\lambda \nu }\mathcal{L}_{F}\right)
\end{equation}%
with $\mathcal{L}_{F}=\frac{\partial \mathcal{L}}{\partial F}.$ One can show
that using $\mathcal{L}\left( F\right) $ given in (12), the Einstein
equations admit the Hayward regular black hole metric provided $\beta =\frac{%
2P^{2}}{\left( 2ml^{2}\right) ^{4/3}}.$ The weak field limit of the
Lagrangian (12) can be found by expanding the Lagrangian about $F=0$ which
reads%
\begin{equation}
\mathcal{L}\left( F\right) =-\frac{6F^{3/2}}{l^{2}\beta ^{3/2}}+\frac{%
12F^{9/4}}{l^{2}\beta ^{9/4}}+\mathcal{O}\left( F^{3}\right) .
\end{equation}%
It is observed that in the weak field limit the NED Lagrangian does not
yield the linear Maxwell Lagrangian i.e., $\lim_{F\rightarrow 0}\mathcal{L}%
\left( F\right) \neq -F.$ For this reason we do not expect that the metric
function in weak field limit gives the RN black hole solution as it was
described in (3).

\section{Stable thin-shell wormhole condition}

In this section we use the standard method of making a timelike TSW and for
this reason to consider a timelike thin-shell located at $r=a$ ($a>r_{h}$)
by cutting the region $r<a$ from the Hayward regular black hole and paste
two copies of it at $r=a$. On the shell the spacetime is chosen to be%
\begin{equation}
ds^{2}=-d\tau ^{2}+a\left( \tau \right) ^{2}\left( d\theta ^{2}+\sin
^{2}\theta d\phi ^{2}\right)
\end{equation}%
in which $\tau $ is the proper time on the shell. To make a consistent $2+1-$%
dimensional timelike shell at the intersection the two $3+1-$dimensional
hypersurfaces we have to fulfill the Lanczos conditions \cite{4}. These are
the Einstein equations on the shell 
\begin{equation}
\left[ K_{i}^{j}\right] -\left[ K\right] \delta _{i}^{j}=-S_{i}^{j}
\end{equation}%
in which a bracket of $X$ is defined as $\left[ X\right] =X_{2}-X_{1},$ $%
K_{i}^{j}$ is the extrinsic curvature tensor in each part of the thin-shell
and $K$ denotes its trace. $S_{i}^{j}$ is the energy momentum tensor on the
shell such that $S_{\tau }^{\tau }=-\sigma $ stands for energy density and $%
S_{\theta }^{\theta }=p=S_{\phi }^{\phi }$ are the surface pressures. One
can explicitly find 
\begin{equation}
\sigma =-\frac{4}{a}\sqrt{f\left( a\right) +\dot{a}^{2}}
\end{equation}%
and%
\begin{equation}
p=2\left( \frac{\sqrt{f\left( a\right) +\dot{a}^{2}}}{a}+\frac{\ddot{a}%
+f^{\prime }\left( a\right) /2}{\sqrt{f\left( a\right) +\dot{a}^{2}}}\right)
.
\end{equation}%
Consequently the energy and pressure densities in a static configuration at $%
a=a_{0}$ are given by%
\begin{equation}
\sigma _{0}=-\frac{4}{a_{0}}\sqrt{f\left( a_{0}\right) }
\end{equation}%
and 
\begin{equation}
p_{0}=2\left( \frac{\sqrt{f\left( a_{0}\right) }}{a_{0}}+\frac{f^{\prime
}\left( a_{0}\right) /2}{\sqrt{f\left( a_{0}\right) }}\right) .
\end{equation}%
To investigate the stability of such a wormhole we apply a linear
perturbation in which the following EoS 
\begin{equation}
p=\psi \left( \sigma \right)
\end{equation}%
with an arbitrary equation $\psi \left( \sigma \right) $ is adopted for the
thin-shell. In addition to this relation between $p$ and $\sigma $ the
energy conservation identity also imposes%
\begin{equation}
S_{\;;j}^{ij}=0
\end{equation}%
which in closed form it amounts to%
\begin{equation}
S_{\;,j}^{ij}+S^{kj}\Gamma _{kj}^{i\mu }+S^{ik}\Gamma _{kj}^{j}=0
\end{equation}%
or equivalently, after the line element (18), 
\begin{equation}
\frac{\partial }{\partial \tau }\left( \sigma a^{2}\right) +p\frac{\partial 
}{\partial \tau }\left( a^{2}\right) =0.
\end{equation}%
This equation can be rewritten as 
\begin{equation}
\dot{a}^{2}+V\left( a\right) =0
\end{equation}%
where $V\left( a\right) $ is given by 
\begin{equation}
V\left( a\right) =f-\left( \frac{a\sigma }{4}\right) ^{4}
\end{equation}%
and $\sigma $ is the energy density after the perturbation. Eq. (28) is a
one dimensional equation of motion in which the oscillatory motion for $a$
in terms of $\tau $ about $a=a_{0}$ is the consequence of having $a=a_{0}$
the equilibrium point which means $V^{\prime }\left( a_{0}\right) =0$ and $%
V^{\prime \prime }\left( a_{0}\right) \geq 0.$ In the sequel we consider $%
f_{1}\left( a_{0}\right) =f_{2}\left( a_{0}\right) $ and therefore at $%
a=a_{0},$ one finds $V_{0}=V_{0}^{\prime }=0.$ To investigate $V^{\prime
\prime }\left( a_{0}\right) \geq 0$ we use the given $p=\psi \left( \sigma
\right) $ to find%
\begin{equation}
\sigma ^{\prime }\left( =\frac{d\sigma }{da}\right) =-\frac{2}{a}\left(
\sigma +\psi \right)
\end{equation}%
and 
\begin{equation}
\sigma ^{\prime \prime }=\frac{2}{a^{2}}\left( \sigma +\psi \right) \left(
3+2\psi ^{\prime }\right) ,
\end{equation}%
with $\psi ^{\prime }=\frac{d\psi }{d\sigma }.$ Finally 
\begin{multline}
V^{\prime \prime }\left( a_{0}\right) = \\
f_{0}^{\prime \prime }-\frac{1}{8}\left[ \left( \sigma _{0}+2p_{0}\right)
^{2}+2\sigma _{0}\left( \sigma _{0}+p_{0}\right) \left( 1+2\psi ^{\prime
}\left( \sigma _{0}\right) \right) \right]
\end{multline}%
where we have used $\psi _{0}=p_{0}.$

\section{Some models of exotic matter supporting the TSW}

Recently two of us analyzed the effect of the Gauss-Bonnet parameter in the
stability of TSW in higher dimensional EGB gravity \cite{10}. In that paper
some specific models of matter have been considered such as LG, CG, GCG,
MGCG and LogG. In this work we go closely to the same EoSs and we analyze
the effect of Hayward's parameter in the stability of the TSW constructed
above.

\subsection{Linear gas (LG)}

\begin{figure}[tbp]
\includegraphics[width=70mm,scale=0.7]{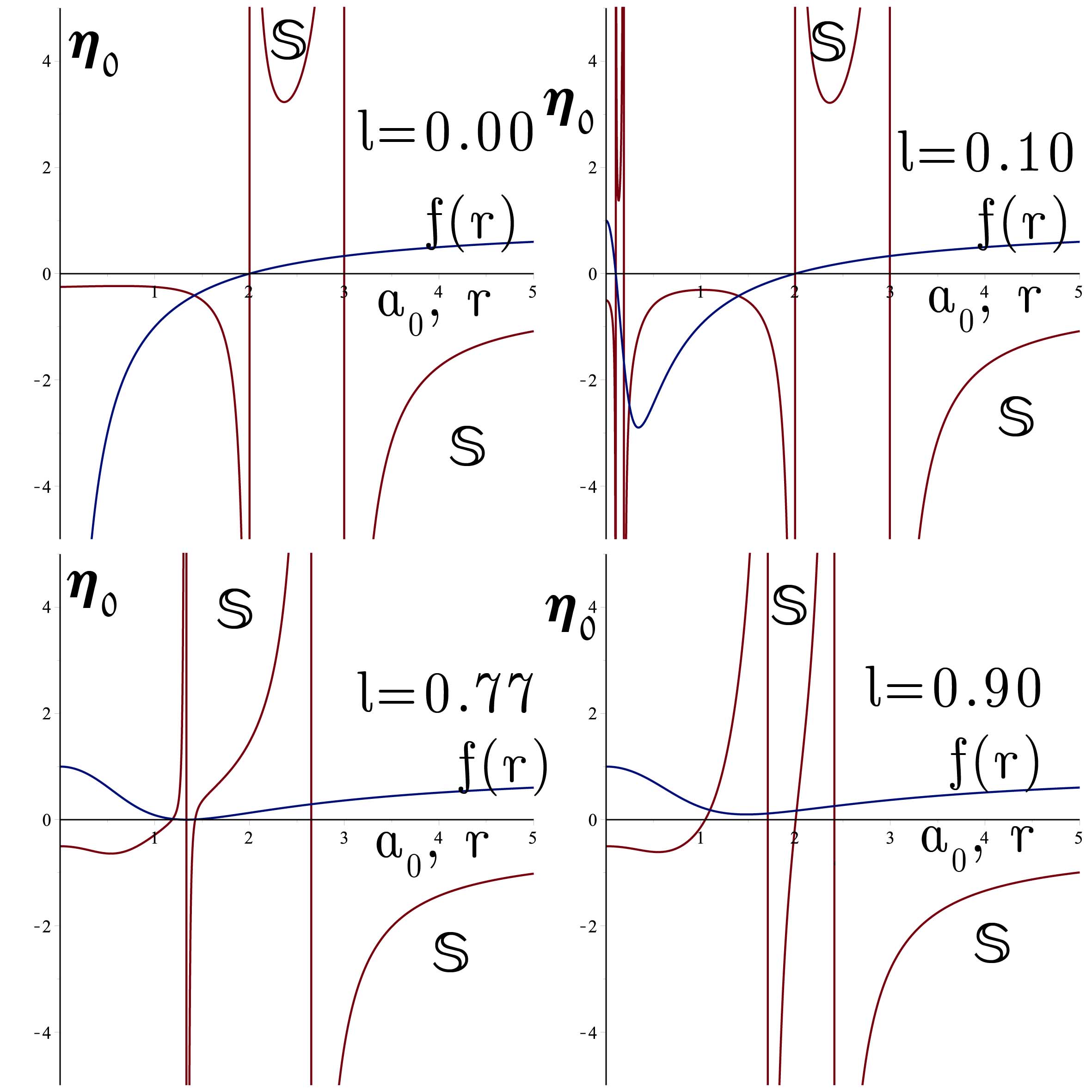}
\caption{Stability of TSW supported by LG in terms of $a_{0}$ and $\protect%
\eta _{0}$ for $\ell =0.00,0.10,0.77$ and $0.90.$ The value of $m=1.$ The
effect of Hayward's constant is to increase the stability of the TSW. We
note that the stable regions are shown by $\mathcal{S}$ and the metric
function is plotted too.}
\end{figure}

In the case of a linear EoS i.e.,%
\begin{equation}
\psi =\eta _{0}\left( \sigma -\sigma _{0}\right) +p_{0}
\end{equation}%
in which $\eta _{0}$ is a constant parameter, one finds $\psi ^{\prime
}\left( \sigma _{0}\right) =\eta _{0}.$ Fig. 1 displays the region of
stability in terms of $\eta _{0}$ and $a_{0}$ for different values of
Hayward's parameter.

\subsection{Chaplygin gas (CG)}

\begin{figure}[tbp]
\includegraphics[width=70mm,scale=0.7]{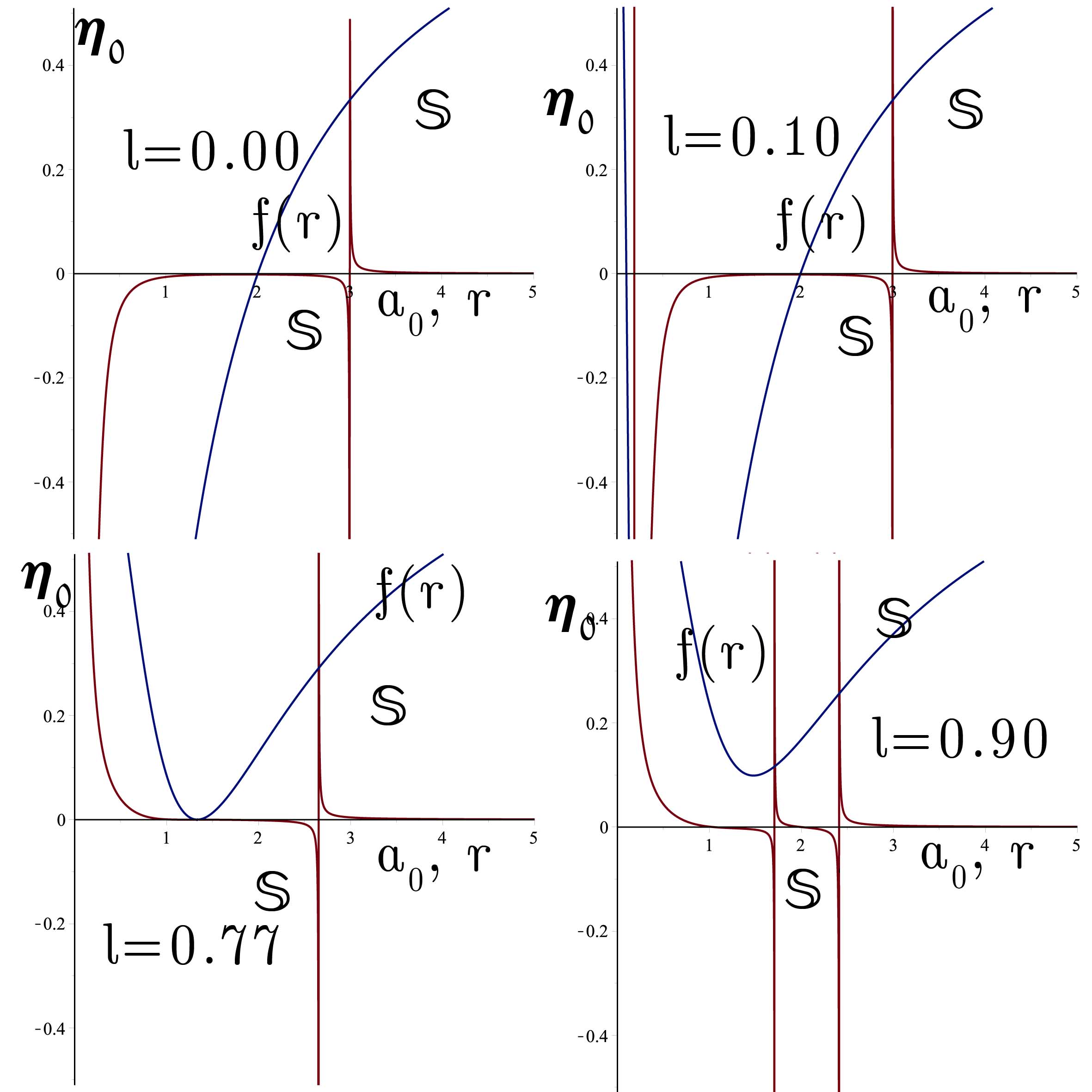}
\caption{Stability of TSW supported by CG in terms of $a_{0}$ and $\protect%
\eta _{0}$ for $\ell =0.00,0.10,0.77$ and $0.90.$ The value of $m=1.$ The
effect of Hayward's constant is to increase the stability of the TSW. We
also plot the metric function to compare the horizon of the black hole and
the location of the throat.}
\end{figure}

For Chaplygin gas (CG) the EoS is given by%
\begin{equation}
\psi =\eta _{0}\left( \frac{1}{\sigma }-\frac{1}{\sigma _{0}}\right) +p_{0}
\end{equation}%
where $\eta _{0}$ is a constant parameter, implies $\psi ^{\prime }\left(
\sigma _{0}\right) =-\frac{\eta _{0}}{\sigma _{0}^{2}}.$ In Fig. 2 we plot
the stability region in terms of $\eta _{0}$ and $a_{0}$ for different
values of $\ell .$

\subsection{Generalized Chaplygin gas (GCG)}

\begin{figure}[tbp]
\includegraphics[width=70mm,scale=0.7]{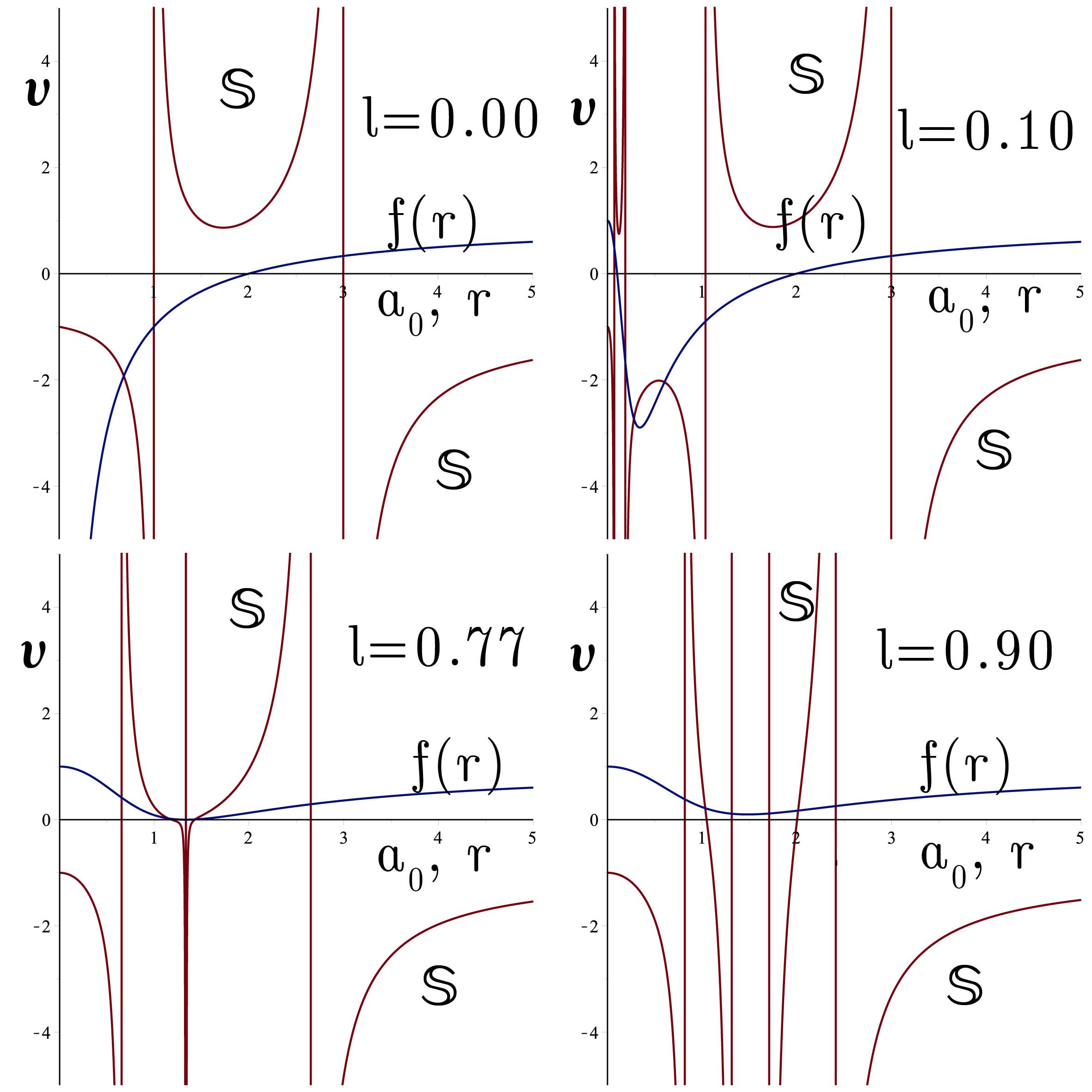}
\caption{Stability of TSW supported by GCG in terms of $a_{0}$ and $\protect%
\nu $ for $\ell =0.00,0.10,0.77$ and $0.90.$ The value of $m=1.$ The effect
of Hayward's constant is to increase the stability of the TSW. We also plot
the metric function to compare the horizon of the black hole and the
location of the throat.}
\end{figure}

The EoS of the Generalized Chaplygin gas can be cast into%
\begin{equation}
\psi \left( \sigma \right) =\eta _{0}\left( \frac{1}{\sigma ^{\nu }}-\frac{1%
}{\sigma _{0}^{\nu }}\right) +p_{0}
\end{equation}%
in which $\nu $ and $\eta _{0}$ are constants. To see the effect of
parameter $\nu $ in the stability we set the constant $\eta _{0}$ such that $%
\psi $ becomes 
\begin{equation}
\psi \left( \sigma \right) =p_{0}\left( \frac{\sigma _{0}}{\sigma }\right)
^{\nu }.
\end{equation}%
We find $\psi ^{\prime }\left( \sigma _{0}\right) =-\frac{p_{0}}{\sigma _{0}}%
\nu $ and in Fig. 3 we plot the stability regions of the TSW supported by a
GCG in terms of $\nu $ and $a_{0}$ with various values of $\ell .$

\subsection{Modified Generalized Chaplygin gas (MGCG)}

\begin{figure}[tbp]
\includegraphics[width=70mm,scale=0.7]{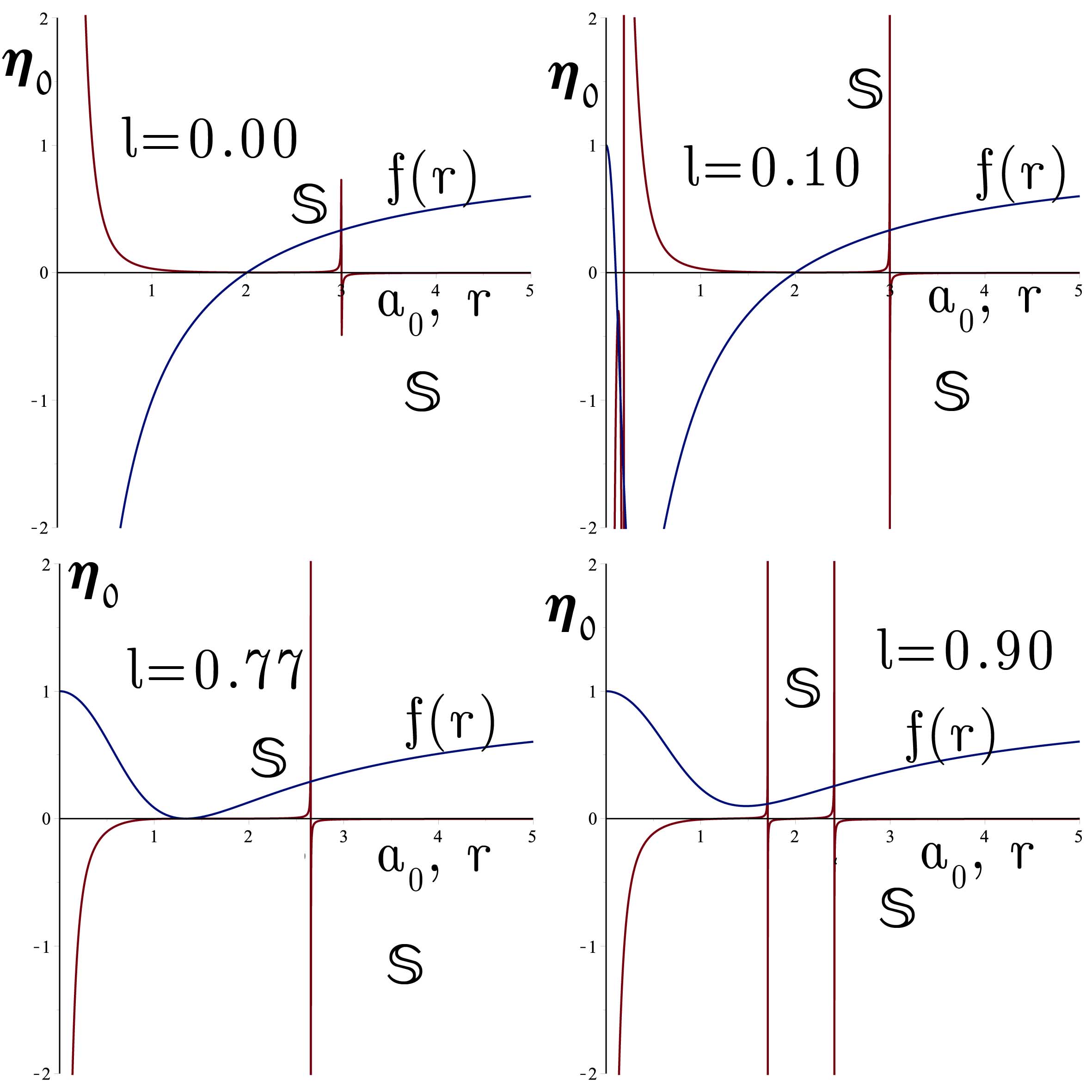}
\caption{Stability of TSW supported by MGCG in terms of $a_{0}$ and $\protect%
\eta _{0}$ for $\ell =0.00,0.10,0.77$ and $0.90.$ The value of $m=1$ and $%
\protect\xi _{0}=\protect\eta _{0}=1.$ The effect of Hayward's constant is
to increase the stability of the TSW. We also plot the metric function to
compare the horizon of the black hole and the location of the throat.}
\end{figure}

A more general form of CG is called the Modified Generalized Chaplygin gas
(MGCG) which is given by%
\begin{equation}
\psi \left( \sigma \right) =\xi _{0}\left( \sigma -\sigma _{0}\right) -\eta
_{0}\left( \frac{1}{\sigma ^{\nu }}-\frac{1}{\sigma _{0}^{\nu }}\right)
+p_{0}
\end{equation}%
in which $\xi _{0}$, $\eta _{0}$ and $\nu $ are free parameters. One then,
finds 
\begin{equation}
\psi ^{\prime }\left( \sigma _{0}\right) =\xi _{0}+\eta _{0}\frac{\eta
_{0}\nu }{\sigma _{0}^{\nu +1}}.
\end{equation}%
To go further we set $\xi _{0}=1$ and $\nu =1$ and in Fig. 4 we show the
stability regions in terms of $\eta _{0}$ and $a_{0}$ with various values of 
$\ell $.

\subsection{Logarithmic gas (LogG)}

\begin{figure}[tbp]
\includegraphics[width=70mm,scale=0.7]{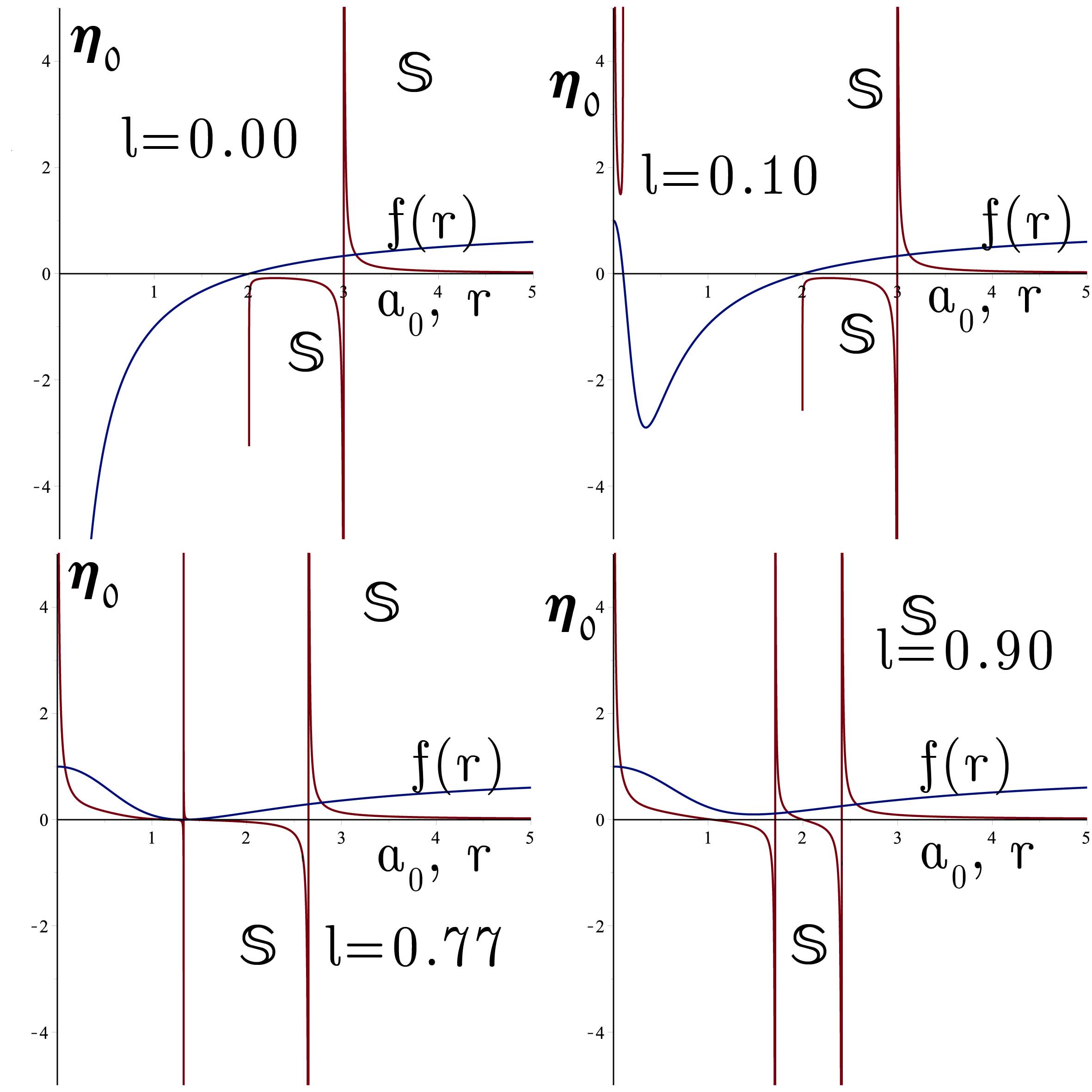}
\caption{Stability of TSW supported by LogG in terms of $a_{0}$ and $\protect%
\eta _{0}$ for $\ell =0.00,0.10,0.77$ and $0.90.$ The value of $m=1.$ The
effect of Hayward's constant is to increase the stability of the TSW. We
also plot the metric function to compare the horizon of the black hole and
the location of the throat.}
\end{figure}

In our last example we consider the Logarithmic gas (LogG) given by%
\begin{equation}
\psi \left( \sigma \right) =\eta _{0}\ln \left\vert \frac{\sigma }{\sigma
_{0}}\right\vert +p_{0}
\end{equation}%
in which $\eta _{0}$ is a constant. For LogG one finds 
\begin{equation}
\psi ^{\prime }\left( \sigma _{0}\right) =\frac{\eta _{0}}{\sigma _{0}}.
\end{equation}%
In Fig. 5 we plot the stability region for the TSW supported by LogG and the
effect of Hayward's parameter is shown clearly.

\section{Stability analysis for small velocity perturbations around the
static solution}

In this section we restrict ourselves to the small velocity perturbations
about the equilibrium point $a=a_{0}$ such that at any proper time after the
perturbation we can consider the fluid supporting the shell to be
approximately static. Thus one can accept the dynamic EoS of the wormhole
same as the static EoS \cite{13}. This assumption, therefore, implies that
the EoS is uniquely determined by $f\left( a\right) $ and $a$, described by
Eqs. (22), (23) i.e.%
\begin{equation}
p=-\frac{1}{2}\left( 1+\frac{af^{\prime }\left( a\right) }{2f\left( a\right) 
}\right) \sigma .
\end{equation}%
With this EoS together with Eqs. (20) and (21) one finds the one-dimentional
motion of the throat given by 
\begin{equation}
\ddot{a}-\frac{f^{\prime }}{2f}\dot{a}^{2}=0.
\end{equation}%
Now, an integration from both sides implies%
\begin{equation}
\dot{a}=\dot{a}_{0}\frac{\sqrt{f}}{\sqrt{f_{0}}}
\end{equation}%
and a second integration gives%
\begin{equation}
\int_{a0}^{a}\frac{da}{\sqrt{f\left( a\right) }}=\frac{\dot{a}_{0}}{\sqrt{%
f_{0}}}\left( \tau -\tau _{0}\right) .
\end{equation}%
Note that $\dot{a}_{0}$ for the equilibrium point is zero but here after
perturbation we assume that the perturbation consists of an initial small
velocity which we call it $\dot{a}_{0}.$

\subsection{The Schwarzschild Example}

The last integral (44) depends on the bulk metric, so that it gives
different results for different spacetimes. For the Schwarzschild bulk, $%
f\left( a\right) =1-\frac{2m}{a}$ which a substitution in (44) yields%
\begin{equation}
\frac{\dot{a}_{0}}{\sqrt{f_{0}}}\left( \tau -\tau _{0}\right) =a\sqrt{f}%
-a_{0}\sqrt{f_{0}}+m\ln \left( \frac{a-m+a\sqrt{f}}{a_{0}-m+a_{0}\sqrt{f_{0}}%
}\right) .
\end{equation}%
This motion is not clearly oscillatory which indicates that the throat is
unstable against the small velocity perturbation.

\subsection{The Hayward Example}

For the case of the Hayward bulk spacetime, Eq. (44), up to the second order
of $\ell ,$ admits 
\begin{multline}
\frac{\dot{a}_{0}}{\sqrt{f_{0}}}\left( \tau -\tau _{0}\right) \tilde{=}a%
\sqrt{f}- \\
a_{0}\sqrt{f_{0}}+m\ln \left( \frac{a-m+a\sqrt{f}}{a_{0}-m+a_{0}\sqrt{f_{0}}}%
\right) + \\
2\ell ^{2}\left( \frac{2a^{2}-2am-m^{2}}{3ma^{2}\sqrt{f}}-\frac{%
2a_{0}^{2}-2a_{0}m-m^{2}}{3ma_{0}^{2}\sqrt{f_{0}}}\right) .
\end{multline}%
Similar to the previous case, this motion is not also oscillatory which
implies that the throat is unstable against the small velocity perturbation.
Nevertheless, Eq. (42) shows that the acceleration of the throat is given by 
$\ddot{a}=\frac{f^{\prime }}{2f}\dot{a}^{2}$ which is positive for both the
Schwarzschild and Hayward bulks. Thus the motion of the throat is not
oscillatory and consequently the corresponding TSW is not stable.

\section{Conclusion}

Thin-shell wormholes are constructed from the regular black hole (or
non-black hole for certain range of parameters) discovered by Hayward. We
show first that this solution is powered by a magnetic monopole field within
the context of non-linear electrodynamics (NED). The non-linear Lagrangian
in the present case can be expressed in a non-polynomial form of the Maxwell
invariant. Such a Lagrangian doesn't admit a linear Maxwell limit. By
employing the spacetime of Hayward and different equations of state of
generic form $p=\psi \left( \sigma \right) $ on the thin-shell we plot
possible stable regions. Amongst these linear, logarithmic and different
Chaplygin gas forms are used and stable regions are displayed. The method of
identifying these regions relies on the reduction of the perturbation
equations to a harmonic equation of the form $\ddot{x}+\frac{1}{2}V^{\prime
\prime }\left( a_{0}\right) x=0$ for $x=a-a_{0}.$ Stability simply amounts
to the condition for $V^{\prime \prime }\left( a_{0}\right) >0$ which is
plotted numerically. In all different equations of state we obtained stable
regions and observed that the Hayward parameter $\ell $ plays a crucial role
in establishing the stability. That is, for higher $\ell $ value we have
enlargement in the stable region. The trivial case $\ell =0$ corresponds to
the Schwarzschild case and is well-known. We have considered also
perturbations with small velocity. It turns out that our TSW is no more
stable against such kind of perturbations. We would like to add here that a
stable spherically symmetric wormhole in general relativity has been
introduced in \cite{11}. Finally, we admit that in each case our energy
density happens to be negative so that we are confronted with exotic matter.
In a separate study we have shown that to have anything but exotic matter to
thread the wormhole we have to abandon spherical symmetry and consider
prolate / oblate spheroidal sources \cite{12}.


\bigskip

\begin{thebibliography}{99}
\bibitem{1} M. Visser, Phys. Rev. D \textbf{39}, 3182 (1989);

M. Visser, Nucl. Phys. \textbf{B} 328, 203 (1989);

P. R. Brady, J. Louko and E. Poisson, Phys. Rev. D \textbf{44}, 1891 (1991);

E. Poisson and M. Visser, Phys. Rev. D \textbf{52}, 7318 (1995);

M. Ishak and K. Lake, Phys. Rev. D \textbf{65}, 044011 (2002);

C. Simeone, Int. Jou. of Mod. Phys. D \textbf{21}, 1250015 (2012);

E. F. Eiroa and C. Simeone, Phys. Rev. D \textbf{82}, 084039 (2010);

F. S. Lobo, Phys. Rev. D \textbf{71}, 124022 (2005);

E. F. Eiroa and C. Simeone, Phys. Rev. D \textbf{71}, 127501 (2005);

E. F. Eiroa, Phys. Rev. D \textbf{78}, 024018 (2008);

F. S. N. Lobo and P. Crawford, Class. Quantum Grav. \textbf{22,} 4869 (2005);

N. M. Garcia, F. S. N. Lobo and M. Visser, Phys. Rev. D \textbf{86}, 044026
(2012);

S. H. Mazharimousavi, M. Halilsoy and Z. Amirabi, Phys. Lett. A \textbf{375}%
, 3649 (2011);

M. Sharif and M. Azam, Eur. Phys. J. C \textbf{73}, 2407 (2013);

M. Sharif and M. Azam, Eur. Phys. J. C \textbf{73}, 2554 (2013);

S. Habib Mazharimousavi and M. Halilsoy, Eur. Phys. J. C \textbf{73}, 2527
(2013).

\bibitem{Dias} G. A. S. Dias and J. P. S. Lemos, Phys. Rev. D \textbf{82},
084023 (2010).

\bibitem{Lobo} J. P. S. Lemos, F. S. N. Lobo, S. Q. Oliveira, Phys. Rev. D 
\textbf{68}, 064004 (2003).

\bibitem{2} J. Bardeen, Proceedings of GR5, Tiflis, U.S.S.R. (1968);

A. Borde , Phys.Rev. D \textbf{50,} 3392(1994);

A. Borde , Phys.Rev. D \textbf{55,} 7615 (1997).

\bibitem{Ayon} E. Ayon-Beato and A. Garc\i a, Phys. Rev. Lett. \textbf{80,}
5056 (1998).

\bibitem{Bronnikov} K. Bronnikov, Phys. Rev. Lett. \textbf{85}, 4641 (2000);

K. Bronnikov, Phys. Rev. D \textbf{63}, 044005 (2001);

K. A. Bronnikov, V. N. Melnikov, G. N. Shikin and K. P. Staniukovich. Ann.
Phys. (USA) \textbf{118}, 84 (1979).

\bibitem{3} S. A. Hayward, Phys. Rev. Lett. \textbf{96}, 031103 (2006).

\bibitem{4} W. Israel, Nuovo Cimento \textbf{44B}, 1 (1966);

V. de la Cruzand W. Israel, Nuovo Cimento \textbf{51A}, 774 (1967);

J. E. Chase, Nuovo Cimento \textbf{67B}, 136. (1970);

S. K. Blau, E. I. Guendelman, and A. H. Guth, Phys. Rev. D \textbf{35}, 1747
(1987);

R. Balbinot and E. Poisson, Phys. Rev. D \textbf{41}, 395 (1990).

\bibitem{5} M. G. Richarte and C. Simeone, Phys. Rev. D \textbf{80}, 104033
(2009);

M. G. Richarte, Phys. Rev. D \textbf{82}, 044021 (2010);

\bibitem{6} E. F. Eiroa and C. Simeone, Phys. Rev. D \textbf{76,} 024021
(2007);

F. S. N. Lobo, Phys. Rev. D \textbf{73}, 064028 (2006);

\bibitem{7} V. Gorini, U. Moschella, A. Y. Kamenshchik, V. Pasquier and A.
A. Starobinsky, Phys. Rev. D \textbf{78,} 064064 (2008);

V. Gorini, A. Y. Kamenshchik, U. Moschella,O. F. Piattella and A. A.
Starobinsky, Phys. Rev. D \textbf{80,} 104038(2009);

E. F. Eiroa, Phys. Rev. D \textbf{80,} 044033 (2009);

C. Bejarano and E. F. Eiroa, Phys. Rev. D \textbf{84}, 064043 (2011);

E. F. Eiroa and G. F. Aguirre, Eur. Phys. J. C \textbf{72}, 2240 (2012).

\bibitem{8} A. Y. Kamenshchik, U. Moschella and V. Pasquier, Phys. Lett. B 
\textbf{487,} 7 (2000);

L. P. Chimento, Phys. Rev. D \textbf{69}, 123517 (2004);

M. Sharif and M. Azam, JCAP \textbf{05}, 25 (2013);

M. Jamil, M. U. Farooq and M. A. Rashid, Eur. Phys. J. C \textbf{59}, 907
(2009).

\bibitem{9} C. Bambi, L. Modesto, Phys. Lett. B \textbf{721,} 329 (2013).

\bibitem{10} Z. Amirabi, M. Halilsoy and S. H. Mazharimousavi, Phys. Rev. D 
\textbf{88}, 124023 (2013).

\bibitem{11} K. A. Bronnikov, L. N. Lipatova, I. D. Novikov and A. A.
Shatskiy, Grav. Cosmol. \textbf{19}, 269 (2013).

\bibitem{12} S. H. Mazharimousavi and M. Halilsoy, "\textit{Thin-shell
wormholes supported by normal matter}": arXiv:1311.6697.

\bibitem{13} M. G. Richarte, Phys. Rev. D \textbf{88}, 027507 (2013);

E. F. Eiroa and C. Simone, Phys. Rev. D \textbf{70}, 044008 (2004);

C. Bejarano, E.\thinspace F. Eiroa, and C. Simeone, Phys. Rev. D \textbf{75}%
, 027501 (2007);

E.\thinspace F. Eiroa and C. Simeone, Phys. Rev. D \textbf{81}, 084022
(2010);

M.\thinspace G. Richarte and C. Simeone, Phys. Rev. D \textbf{79}, 127502
(2009);

E. Rub\'{\i}n de Celis, O.\thinspace P. Santill\'{a}n, and C. Simeone, Phys.
Rev. D \textbf{86}, 124009 (2012);

M. Sharif and M. Azam, J. Cosmol. Astropart. Phys. \textbf{04}, 023 (2013).
\end{thebibliography}
\end{document}